\documentclass[12pt]{article}
\usepackage{amsxtra,amssymb,amsthm,amsmath,latexsym}

\theoremstyle{plain}

\newtheorem{theorem}{Theorem}[section]

\def\oH{\buildrel\circ\over H}
\def\oH1{\buildrel\circ\over H\kern-.02in{}^1}

\begin{document}


\title{Regularization of Ill-Posed Problems with Unbounded Operators.
   \thanks{key words:regularization, ill posed problems, unbounded operators
    }
   \thanks{Math subject classification: 47A52 65J 20}
}

\author{
A.G. Ramm\\
 Mathematics Department, Kansas State University, \\
 Manhattan, KS 66506-2602, USA\\
ramm@math.ksu.edu\\
}

\date{}

\maketitle\thispagestyle{empty}

\begin{abstract}
Variational regularization and the quasisolutions method are justified for
unbounded closed operators.

\end{abstract}


\section{Introduction. Variational regularization.}
There is a large literature on methods for solving ill-posed problems:
variational regularization, method of quasisolutions, iterative and
projection regularization \cite{2}-\cite{8}. The case of 
ill-posed
problems with a closed linear operator was discussed
in [5], and the case of nonlinear, possibly unbounded, operators 
possibly unbounded, does not seem to be discussed. 
In the theory of ill-posed problems the following well-known
result ([1, Lemma I.5.8]) is often used: if $A$ is an injective and 
continuous mapping
from a compact set $M$ of a Banach space into a set $N:=AM$,
then the inverse mapping $A^{-1}: N\to M$ is continuous.
In [7, p.112] the usual assumption
about continuity of $A$ in the above result  is replaced by the assumption 
about closedness of $A$.

In this short note ill-posed problems are studied in the case of
the mapping $A$ not necessarily continuous,
but closed, possibly nonlinear. Our argument is very simple and
the result is fairly general. 

Let $A$ be an injective, possibly nonlinear, closed operator on a Banach
space $X,$ and the equation
$$Ay=f \eqno{(1.1)}$$
has a solution $y$. Our arguments hold in metric spaces as well
without changes.

Assume that $A^{-1}$ is not continuous.
This implies that problem (1.1) is ill-posed. Let
$$\| f_\delta - f \| \leq \delta. \eqno{(1.2)}$$
Given $f_\delta$ and $A$, one wants to construct
$u_\delta = R_\delta (f_\delta)$, such that
$\| u_\delta - y \| \to 0$ as $\delta \to 0$, where $y$ solves (1.1). If
$u_\delta$ is constructed, then the operator $R_\delta$ yields a stable
approximation of the solution $y$ to (1.1).

Let us first describe the method of variational regularization in our case.

Define the functional
$$F(u) := \| A(u) - f_\delta \| + \delta \phi (u), \eqno{(1.3)}$$
and assume that $\phi (u)\geq 0$ is a functional, such that
for any constant $c>0$ the set
$$\{u : \phi (u) \leq c\} \hbox{\ is\ } \hbox{precompact\ } \hbox{in\ } X.
  \eqno{(1.4)}$$
The functional $F$ depends on $\delta$ and $f_\delta$, but
for simplicity of writing we do not
show this dependence explicitly.
Assume that $D(A)\subset D(\phi)$, the domain of definition of $\phi$, 
contains $D(A)$. This assumption implies that $y \in D(\phi)$,
so that $\phi (y)<\infty$. Define $D(F) = D(A)$.
If $A$ were bounded, defined on all of $X$, then one would assume
$y\in  D(\phi)$ and $D(F) = D(\phi)$. If  $A$ were unbounded and 
$ D(\phi)\subset  D(A)$, then one would assume that $y\in  D(\phi)$ and 
$D(F) = 
D(\phi)$.
 
Denote
$$
0 \leq m := \inf_{u \in D(A)} F(u). \eqno{(1.5)}
$$
The number $m=m(\delta)\geq 0$.
Let $u_j$ be a minimizing sequence $u_j \in D(F)$ for the functional
$F$, such that:
$$F(u_j) \leq m + \frac{1}{j} = m + \delta, \quad \frac{1}{j} \leq \delta.
  \eqno{(1.6)}$$
Denote by $u_\delta:=u_{j(\delta)}$ a member $u_{j(\delta)}$
of this minimizing sequence, where
$j(\delta)$ is chosen so that
$\frac{1}{j(\delta)} \leq \delta$. There are many such $j(\delta)$ and
we fix one of them, for example, the minimal one. Since
$$F(y) \leq \delta + \delta \phi (y) := c_1 \delta,
  \quad c_1 := 1+ \phi (y),
  \eqno{(1.7)}$$
one has:
$$m \leq c_1 \delta, \eqno{(1.8)}$$
and
$$F(u_\delta) \leq m + \delta \leq c \delta, \quad c := c_1 + 1.
  \eqno{(1.9)}$$
Thus
$\delta \phi (u_\delta) \leq c \delta$, and
$$\phi (u_\delta) \leq c. \eqno{(1.10)}$$

Let us now take $\delta\to 0$.
By (1.4) and (1.10) one can select a convergent in $X$ subsequence
of the set $u_\delta$, which
we denote also $u_\delta$, such that
$$\| u_\delta - u \| \to 0 \hbox{\ as\ } \delta \to 0, \eqno{(1.11)}$$
where $u$ is the limit of $u_\delta$.

From (1.2), (1.3), (1.9) and (1.10), one gets
$$0 = \lim_{\delta \to 0} F(u_\delta) = \lim_{\delta \to 0}
  \| A(u_\delta) - f_\delta \| = \lim_{\delta \to 0}
  \| A(u_\delta) - f \|. \eqno{(1.12)}$$
Since $A$ is closed, (1.11) and (1.12) imply
$$\lim_{\delta \to 0} A(u_\delta) = A(u), \quad 0 = \|A(u) - f \|.
  \eqno{(1.13)}$$
Since $A$ is injective, (1.13) and (1.1) imply $u=y$, so
$$\lim_{\delta \to 0} \| u_\delta - y \| = 0. \eqno{(1.14)}$$
Since the limit $y$ of any subsequence $u_\delta$ is unique, the
whole sequence $u_\delta$ converges to $y$.

We have proved the following result:
\begin{theorem} 
Assume that (1.4) holds, $\phi\geq 0$,  $A:D(A) \to X$ is a closed,
injective, possibly nonlinear unbounded operator, $A(y) = f$,
and $A^{-1}$ is not continuous. Let 
$u_\delta$
be constructed as above so that (1.9)
holds. Then (1.14) holds.
\end{theorem}

In section 2 the method of quasisolutions is discussed in the case of
possibly unbounded and nonlinear operators.

\section{Quasisolutions for unbounded operators.}
In this section the assumptions about equation (1.1) and the operator $A$ are
the same as in section 1, in particular, $A^{-1}$ is not continuous,
so that solving equation (1.1) is an ill-posed problem.

Choose a compactum $K \subset X$ such that the solution of (1.1)
$y \in K$. Consider the problem
$$\| A(u) - f_\delta \| = \inf := \mu, \quad u \in K \subset D(A).
  \eqno{(2.1)}$$
The infimum $\mu=\mu(\delta)\geq 0$ depends on $f_\delta$ also,
but we do not show this dependence explicitly.
Let $u_j$ be a minimizing sequence:
$$\| A(u_j) - f_\delta \| \leq \mu + \frac{1}{j}. \eqno{(2.2)}$$
Choose $j=j(\delta)$ such that $\frac{1}{j} \leq \delta$ and denote
$u_j := u_\delta$.

Then
$$\| A(u_\delta) - f_\delta \| \leq \mu + \delta. \eqno{(2.3)}$$
Since $\| A(y) - f_\delta \| \leq \delta$, it follows that $\mu 
\leq \delta$, so
$$\| A(u_\delta) - f_\delta \| \leq 2\delta. \eqno{(2.4)}$$
Since $\{u_\delta\} \subset K$ one can select a convergent (to some $u$)
subsequence, denoted also $\{u_\delta\}$:
$$\| u_\delta - u \| \to \hbox{\ as\ } \delta \to 0. \eqno{(2.5)}$$
From (2.4) it follows that $A(u_\delta)$ converges to $f$:
$$\| A(u_\delta) - f \| \leq \| A(u_\delta) - f_\delta \| +
  \| f_\delta - f \| \leq 3 \delta \to 0 \hbox{\ as\ } \delta \to 0.
  \eqno{(2.6)}$$
Since $A$ is closed, it follows from (2.5) and (2.6) that
$$A(u)=f. \eqno{(2.7)}$$
Injectivity of $A$, equation (2.7), and the equation $A(y)=f$ imply $u=y$.
We have proved:

\begin{theorem} 
Assume that $A:D(A) \to X$ is a closed, 
injective, possibly nonlinear and unbounded, operator, $A(y) = f$, 
(1.2) holds, $K$ is a compact set in $X$, and $y\in K$.
If $u_\delta \in K$  satisfies (2.4), then (1.14) holds.
\end{theorem}

{\bf Acknowledgement.} The author thanks Professor A.Yagola
for useful discussions.

\end{document}